\documentclass[12pt]{book}

\usepackage[dvips]{graphicx,color}
\usepackage{makeidx,physics,cosmology}
\usepackage{latexsym}
\usepackage{amsfonts}
\usepackage{amssymb}
\usepackage{amsmath}

\makeauthorindex
\makeindex

\BookTitle{Frontier in Astroparticle Physics and Cosmology}
\CopyRight{\copyright 2004 by Universal Academy Press, Inc.}

\begin{document}

\BookTitle{\itshape Frontier in Astroparticle Physics and Cosmology}
\CopyRight{\copyright 2004 by Universal Academy Press, Inc.}
\pagenumbering{arabic}

\chapter{
Inflationary Braneworld Driven By Bulk Scalar Field}

\author{%
Jiro SODA\\
{\it Department of Physics, Kyoto University, Kyoto 606-8501, Japan}\\
Sugumi KANNO\\
{\it Department of Physics, Kyoto University, Kyoto 606-8501, Japan}}
%
%
\AuthorContents{J.\ Soda and S.\ Kanno} 

\AuthorIndex{Soda}{J.} 
\AuthorIndex{Kanno}{S.}

\section*{Abstract}
 We have developed a formalism to
 study an inflationary scenario driven by a bulk inflaton in the two-brane
 system. The 4-dimensional low energy effective action is obtained 
 using the gradient expansion method. 
 It is also found that the dark radiation and the dark 
 scalar source are expressed by the radion. In the single-brane limit,
 we find these dark components disappear. Therefore, it turns out that
 the inflation due to the bulk inflaton  successfully takes place. 
 Kaluza-Klein corrections are also taken into account in this case. 

\section{Introduction}

  Inspired by the recent 
 developments of the superstring theory, Randall and Sundrum have proposed
 a novel compactification mechanism in the context of the brane world 
 scenario~\cite{RS1}. It is natural to ask if the inflationary universe
 can be possible or not in this context. 
 Let us recall the well known formula for the cosmological constant
in the braneworld: 
\begin{eqnarray}
    \Lambda_{\rm eff} = {\kappa^4 \sigma^2  \over 12}
                      - {3 \over \ell^2 } \ ,
\end{eqnarray}
where $\sigma$ and $\ell$ are the tension of the brane and the curvature
scale in the bulk which is determined by the bulk vacuum energy, respectively.
 For $\kappa^2 \sigma = 6/\ell$, we have Minkowsky spacetime.
 In order to obtain the inflationary universe, we need the positive
 effective cosmological constant. In the brane world model, there are
 two possibilities. One is to increase the brane tension and the other
 is to increase $\ell$. The brane tension can be controlled by the scalar
 field on the brane. The bulk curvature scale $\ell$ can be controlled
  by the bulk scalar field. The former case is a natural extension of the
  4-dimensional inflationary scenario. The latter possibility
  is a novel one peculiar to the brane model. 
   Recall that, in the superstring theory,  scalar fields  are 
 ubiquitous. Indeed, the dilaton and moduli exists in the bulk 
 generically, because they arise as the  modes associated with the 
 closed string. Moreover, when the supersymmetry is spontaneously broken,
 they may have the non-trivial potential. Hence, it is natural to consider 
 the inflationary scenario driven by these 
 fields~\cite{bulk}. 
 In this paper, we would like to present 
 a formalism to discuss the inflationary scenario in the braneworld context. 
 We also show the bulk inflaton can, in fact, drive the 
 inflation on the brane. 
  
 The organization of this paper is as follows.
 In sec.2, we describe the model. In sec.3, we obtain the effective action
 using the low energy approximation and discuss its implication. 
 In sec. 4, the action with KK effects are presented. 
 In the final section, we summarize our  results.

\section{Model and Basic Equations}
 
 We consider a $S_1/Z_2$ orbifold spacetime with the two branes 
as the fixed points. In this first Randall-Sundrum (RS1) model, 
the two flat 3-branes are embedded in AdS$_5$ 
 and the brane tensions given by
$\overset{\oplus}{\sigma}=6/(\kappa^2\ell)$ and 
$\overset{\ominus}{\sigma}=-6/(\kappa^2\ell)$. 
 Our system is described by the action 
\begin{eqnarray}
S &=& {1\over 2\kappa^2}\int d^5 x \sqrt{-g}~{\cal R}
	-\int d^5 x  \sqrt{-g} \left[~ 
	{1\over 2} g^{AB} \partial_A\varphi 
	\partial_B\varphi
	+ U(\varphi) ~\right]\nonumber\\
&&	-\sum_{i=\oplus,\ominus} \overset{i}{\sigma} 
\int d^4 x \sqrt{-g^{i\mathrm{\hbox{-}brane}}} 
+\sum_{i=\oplus,\ominus} \int d^4 x \sqrt{-g^{i\mathrm{\hbox{-}brane}}}
\,{\cal L}_{\rm matter}^i  \ ,
\label{5D:action}
\end{eqnarray}
where  $g^{i\mathrm{\hbox{-}brane}}_{\mu\nu}$ 
and $\overset{i}{\sigma}$ are  the induced metric 
 and the brane tension on the $i$-brane, respectively. 
We assume the potential $U(\varphi)$ for the bulk scalar field takes the form
$
U(\varphi)=-\frac{6}{\kappa^2\ell^2}+V(\varphi)\ ,
\label{potential}
$
where the first term is regarded as a 5-dimensional cosmological constant
and the second term is an arbitrary potential function.
The brane tension $\sigma$ is tuned so that
the effective cosmological constant on the brane vanishes. 
The above setup realizes a flat braneworld after inflation ends and 
the field $\varphi$ reaches the minimum of its potential. 

Inflation in the braneworld can be driven by a scalar field 
either on the brane or in the bulk.
We derive the effective equations of motion which are useful for both 
models. 
In this section, we begin with the single-brane system. Since we 
know the effective 4-dimensional equations hold irrespective of the existence
 of other branes~\cite{kanno1},
  the analysis of the single-brane system is sufficient
 to derive the effective action for the two-brane system 
 as we see in the next section.

We adopt the Gaussian normal coordinate system to describe 
the geometry of the brane model;
$
ds^2 = dy^2 + g_{\mu\nu} (y,x^{\mu} ) dx^{\mu} dx^{\nu}  \ ,
$
where the brane is assumed to be located at $y=0$. 
 Let us decompose the extrinsic curvature into the traceless part 
 $\Sigma_{\mu\nu}$ and the trace part $K$ as
$
K_{\mu\nu}=-\frac{1}{2}g_{\mu\nu,y}
	=\Sigma_{\mu\nu}
	+{1\over 4} g_{\mu\nu} K \  .
$
Then, we can obtain the basic equations off the brane 
using these variables.
First, the Hamiltonian constraint equation leads to
\begin{eqnarray}       		
{3\over 4} K^2-\Sigma^{\alpha}{}_{\beta}\Sigma^{\beta}{}_{\alpha}
	&=& \overset{(4)}{R} -\kappa^2 \nabla^\alpha\varphi
	\nabla_\alpha\varphi 
	+\kappa^2(\partial_y\varphi )^2
	-2\kappa^2U(\varphi ) \ ,  
    	\label{eq:hamiltonian} 
\end{eqnarray}
where $\overset{(4)}{R}$ is the curvature on the brane and 
$\nabla_\mu $ denotes the
covariant derivative with respect to the metric $g_{\mu\nu}$. Momentum 
constraint equation becomes
\begin{eqnarray}       		
\nabla_\lambda \Sigma_{\mu}{}^{\lambda}  
	-{3\over 4} \nabla_\mu K =-\kappa^2
	\partial_y\varphi\partial_\mu\varphi\ .
	\label{eq:momentum}  
\end{eqnarray}
Evolution equation in the direction of $y$ is given by
\begin{eqnarray}
\Sigma^{\mu}{}_{\nu,y}-K\Sigma^{\mu}{}_{\nu} 
	=-\left[ \overset{(4)}{R}{}_{\mu\nu}-\kappa^2 \nabla^\mu\varphi
	\nabla_\nu\varphi 
	\right]_{\rm traceless}      \ .     
	\label{eq:evolution} 
\end{eqnarray}
Finally,  the equation of motion for the scalar field reads
\begin{eqnarray}
\partial^2_y \varphi -K\partial_y\varphi 
	+ \nabla^{\alpha}\nabla_{\alpha}\varphi -U'(\varphi)=0  \ ,
	\label{eq:scalar}	
\end{eqnarray}
where the prime denotes derivative with respect to the scalar field $\varphi$.

 As we have the singular source at the brane position, we must consider 
 the junction conditions. 
 Assuming a $Z_2$ symmetry of spacetime, we obtain the junction 
conditions for the metric and the scalar field 
\begin{eqnarray}
\left[ \Sigma^{\mu}{}_{\nu}-{3\over 4} \delta^\mu_\nu K \right] \Bigg|_{y=0}
	&=& -{\kappa^2 \over 2}\sigma  \delta^\mu_\nu 
    	+\frac{\kappa^2}{2}T^{\mu}{}_{\nu}  \ , 
    	\label{JC:metric} \\
	\Bigl[\partial_y\varphi\Bigr]\bigg|_{y=0} &=& 0 \ ,
	\label{JC:scalar}         
\end{eqnarray}
where $T^{\mu}{}_{\nu}$ is the energy-momentum tensor for the matter
fields  on the brane.

\section{Low Energy Effective Action}

 We assume the inflation occurs at low energy
 in the sense that the additional energy due to the bulk scalar field is small, 
 $\kappa^2\ell^2V(\varphi)\ll 1$, and the curvature on the brane $R$ 
 is also small, $R\ell^2\ll 1$.  It should be stressed that the low energy
 does not necessarily implies weak gravity on the brane. 
 Under these circumstances, we can use a
 gradient expansion scheme to solve the bulk equations of motion.  

 At zeroth order,  we ignore matters on the brane.
  Then, from the junction condition (\ref{JC:metric}), we have 
\begin{eqnarray} 
\left[ \overset{(0)}{\Sigma}{}^{\mu}{}_{\nu}
 -{3\over 4} \delta^\mu_\nu \overset{(0)}{K} \right] \Bigg|_{y=0}
	&=& -{\kappa^2 \over 2}\sigma \delta^\mu_\nu  \ .
	\label{JC:metric0}
\end{eqnarray}
 As the right hand side of (\ref{JC:metric0}) contains no traceless part,
 we get 
$
\overset{(0)}{\Sigma}{}^\mu{}_\nu  =0 \ .
$
We also take the potential for the bulk scalar field 
$U(\varphi)$ to be $-6/(\kappa^2\ell^2)$. 
We discard the terms with  4-dimensional derivatives since one can neglect
 the long wavelength variation in the direction of $x^\mu$ at low energies.  
 Thus, the equations to be solved are given by
\begin{eqnarray}       		
  && {3\over 4} \overset{(0)}{K}{}^2
	=  	  \kappa^2(\partial_y \overset{(0)}{\varphi} )^2
	  + {12 \over \ell^2 } \ ,  \\
    	\label{eq:hamiltonian0} 
  && \partial^2_y \overset{(0)}{\varphi} 
        - \overset{(0)}{K} \partial_y  \overset{(0)}{\varphi} =0  \ .
	\label{eq:scalar0}	
\end{eqnarray}
The  junction condition (\ref{JC:scalar}) at this order
$
  \Bigl[\partial_y \overset{(0)}{\varphi} \Bigr]\bigg|_{y=0} = 0
$
tells us that the solution of Eq.(\ref{eq:scalar0}) must be
$
\overset{(0)}{\varphi}=\eta~(x^\mu) \ ,
$
where $\eta (x^\mu )$ is an arbitrary constant of integration.
 Now, the solution of  Eq.(\ref{eq:hamiltonian0}) yields
$
\overset{(0)}{K}= {4 / \ell} \ .
$
Other Eqs. (\ref{eq:momentum}) and (\ref{eq:evolution})
 are trivially satisfied at zeroth order. 
Using the definition
$\overset{(0)}{K}{}_{\mu\nu}= -\overset{(0)}{g}_{\mu\nu,y}/2$,  
we have the lowest order metric 
\begin{equation}
\overset{(0)}{g}{}_{\mu\nu} (y,x^\mu ) 
  = b^2(y)~h_{\mu\nu} (x^\mu ), \qquad 
b(y)\equiv e^{- y/ \ell} \ ,
\label{formal:metric}
\end{equation}
where the induced metric on the brane, $ h_{\mu\nu}\equiv g_{\mu\nu} 
(y=0 ,x^\mu) $, arises as a constant of  integration. 
The junction condition for the induced metric (\ref{JC:metric0}) merely 
implies well known relation $\kappa^2\sigma =6/\ell$
and that for the scalar field (\ref{JC:scalar}) is trivially satisfied.
At this leading order analysis, we can not determine the constants of 
integration $h_{\mu\nu} (x^\mu)$ and $\eta (x^\mu)$ which are constant
 as far as the short length scale $\ell$ variations are concerned, but
 are allowed to vary over the long wavelength scale. These constants
 should be  constrained by the next order analysis. 

 Now, we take into account the effect of both the bulk scalar field 
and the matter on the brane perturbatively.
Our iteration scheme 
is to write the metric $g_{\mu\nu}$ and the scalar field $\varphi$ 
as a sum of local tensors built out of the  induced metric and the induced 
 scalar field on the brane,  in the order of expansion parameters, that is, 
$O((R \ell^2 )^{n})$ and $O(\kappa^2\ell^2V(\varphi))^n$, 
$n=0,1,2,\cdots$~\cite{kanno1}. 
 Then, we expand the metric and the scalar field  as 
\begin{eqnarray}
&& g_{\mu\nu} (y,x^\mu ) =
	b^2 (y) \left[ h_{\mu\nu} (x^\mu) 
  	+ \overset{(1)}{g}_{\mu\nu} (y,x^\mu)
      	+ \overset{(2)}{g}_{\mu\nu} (y, x^\mu ) + \cdots  \right]  \ , 
      	\label{expansion:metric} \nonumber\\
&& \varphi(y,x^\mu) = \eta (x^\mu) 
	+ \overset{(1)}{\varphi} (y, x^\mu) +\overset{(2)}{\varphi} (y, x^\mu)
	+ \cdots  \ .
\end{eqnarray}
Here, we put the boundary conditions
 $ \overset{(i)}{g}_{\mu\nu} (y=0 ,x^\mu ) =  0  
	\ ,   \overset{(i)}{\varphi } (y=0 ,x^\mu) =0 \ ,\quad i=1,2,3,...
$
so that we can interpret $h_{\mu\nu}$ and $\eta$ as induced quantities. 
Extrinsic curvatures can be also expanded as 
\begin{eqnarray}
K = \frac{4}{\ell} 
	+\overset{(1)}{K}
	+\overset{(2)}{K}
	+\cdots  \ , \qquad
\Sigma^\mu_{\ \nu} = 
	\overset{(1)}{\Sigma}{}^{\mu}_{\ \nu}
        +\overset{(2)}{\Sigma}{}^{\mu}_{\ \nu}
        +\cdots \ .
        \label{expansion:sigma}
\end{eqnarray}
 Using the formula such as 
$\overset{(4)}{R} (\overset{(0)}{g}_{\mu\nu})=R(h_{\mu\nu})/b^2$, 
we obtain the solution
\begin{eqnarray}
\overset{(1)}{K}=\frac{\ell}{6b^2}\left(
	R (h) -\kappa^2\eta^{|\alpha}\eta_{|\alpha}\right)
	-\frac{\ell}{3}\kappa^2V(\eta) \ ,
	\label{1:hamiltonian} 
\end{eqnarray}
where $R(h)$ is the scalar curvature of $h_{\mu\nu}$
 and $|$ denotes the covariant derivative with respect to  
$h_{\mu\nu}$. 
 Substituting the results at zeroth order solutions into 
Eq.~(\ref{eq:evolution}), we obtain
\begin{eqnarray}
\overset{(1)}{\Sigma}{}^{\mu}{}_{\nu}=
	\frac{\ell}{2b^2}\left[
	R^{\mu}{}_\nu (h) -\kappa^2\eta^{|\mu}\eta_{|\nu}\right]_{\rm traceless}
	+\frac{\chi^{\mu}{}_\nu}{b^4} \ ,
	\label{1:evolution}
\end{eqnarray}
where $R^{\mu}{}_\nu (h)$ denotes the Ricci tensor of $h_{\mu\nu}$ and 
$\chi^{\mu}{}_\nu$ is a constant of integration which 
satisfies the constraint $\chi^{\mu}{}_{\mu}=0$. 
 Hereafter, we omit the argument of the curvature for simplicity. 
 Integrating the scalar field equation (\ref{eq:scalar}) at first order, 
we have
\begin{eqnarray}
\partial_y\overset{(1)}{\varphi}&=&\frac{\ell}{2b^2} \Box \eta
	-\frac{\ell}{4}V'(\eta)
	+\frac{C}{b^4} \ ,
	\label{1:scalar}
\end{eqnarray}
where $C$ is also a constant of integration. 
At first order in this iteration scheme,
we get two kinds of constants of integration, $\chi^{\mu}{}_{\nu}$ and $C$.

Given  the matter fields $T_{\mu\nu}$ on the brane,  the junction 
condition (\ref{JC:metric}) becomes
\begin{eqnarray}
\left[\overset{(1)}{\Sigma}{}^{\mu}{}_{\nu}
	-\frac{3}{4}\delta^\mu_\nu
	\overset{(1)}{K}\right] \Bigg|_{y=0} 
	=    \frac{\kappa^2}{2}T^{\mu}{}_{\nu} \ .
	\label{1:JC-m}
\end{eqnarray}
 At this order, the junction condition (\ref{JC:scalar}) yields
\begin{eqnarray}
\biggl[\partial_y\overset{(1)}{\varphi}\biggr] \Bigg|_{y=0}
	=  0  \ .
	\label{1:JC-s}
\end{eqnarray}
These junction conditions give the effective equations of motion
 on the brane.

The point is the fact that the equations of motion on each brane
take the same form if we use the induced metric on each brane~\cite{kanno2}. 
 The effective Einstein equations on each positive ($\oplus$)
and negative ($\ominus$) tension brane at low-energies yield
\begin{eqnarray}
G^\mu{}_\nu (h) &=& \kappa^2 \left( 
	\eta^{|\mu}\eta_{|\nu}
        -{1\over 2}\delta^\mu_\nu \eta^{|\alpha} \eta_{|\alpha}
        -{1\over 2}\delta^\mu_\nu V\right) 
        -{2\over \ell} \chi^\mu{}_\nu
        +\frac{\kappa^2}{\ell}  
        \overset{\oplus}{T}{}^{\mu}{}_{\nu}
        \label{p:einstein}    \ , \\
G^\mu{}_\nu (f) &=& \kappa^2 \left( 
        \eta^{;\mu} \eta_{;\nu}
        -{1\over 2}\delta^\mu_\nu \eta^{;\alpha} \eta_{;\alpha}
        -{1\over 2}\delta^\mu_\nu V\right)
        -{2\over \ell} {\chi^\mu{}_\nu \over \Omega^4}
        -\frac{\kappa^2}{\ell} 
        \overset{\ominus}{T}{}^{\mu}{}_{\nu} 
        \label{n:einstein-1}  \ .
\end{eqnarray}
where $f_{\mu\nu}$ is the induced metric on the negative tension brane 
and $;$ denotes the covariant derivative with respect to $f_{\mu\nu}$. 
 When we set the position of the positive tension brane at $y=0$, 
  that of the negative tension brane $\bar{y}$ in general depends on
 $x^\mu$, i.e. $\bar{y} = \bar{y} (x^\mu)$. Hence,  
 the warp factor at the negative tension brane 
 $\Omega (x^\mu)\equiv b(\bar{y}(x)) $ also depends on $x^\mu$.
  Because the metric always comes into equations
 with  derivatives, the zeroth order relation is enough in this 
 first order discussion. Hence, the metric on the
 positive tension brane is related to the metric on the 
 negative tension brane as $f_{\mu\nu} = \Omega^2 h_{\mu\nu}$.
   
Although Eqs.~(\ref{p:einstein}) and (\ref{n:einstein-1}) are non-local 
individually, with undetermined $\chi^{\mu}{}_{\nu}$, one can combine both
equations to reduce them to local equations for each brane. 
We can therefore easily eliminate $\chi^{\mu}{}_{\nu}$ from 
Eqs.~(\ref{p:einstein}) and (\ref{n:einstein-1}), since 
$\chi^{\mu}{}_{\nu}$ appears only algebraically.
Eliminating $\chi^\mu{}_\nu$ from both Eqs.~(\ref{p:einstein}) and 
(\ref{n:einstein-1}), we obtain 
\begin{eqnarray}
G^\mu{}_\nu &=&
	\frac{\kappa^2}{\ell\Psi}
	\overset{\oplus}{T}{}^{\mu}{}_{\nu}
	+\frac{\kappa^2(1-\Psi)^2}{\ell\Psi}
	\overset{\ominus}{T}{}^{\mu}{}_{\nu}
        \nonumber\\
&&
	+{1\over \Psi}\left[ \Psi^{|\mu}{}_{|\nu} 
     	- \delta^\mu_\nu \Psi^{|\alpha}{}_{|\alpha} 
     	+{3\over 2}{1 \over 1-\Psi} \left(
     	\Psi^{|\mu} \Psi_{|\nu} -{1\over 2}\delta^\mu_\nu
     	\Psi^{|\alpha} \Psi_{|\alpha} \right) \right] \nonumber\\
&& 
	+\kappa^2\left( 
        \eta^{|\mu} \eta_{|\nu}
        -\frac{1}{2}\delta^\mu_\nu\eta^{|\alpha}\eta_{|\alpha}
	-\delta^\mu_\nu V_{\rm eff}
        \right)
        \label{QST1}
         \ ,  \quad V_{\rm eff}=  \frac{2-\Psi}{2}V \ ,
\end{eqnarray}
where we defined a new field $\Psi=1-\Omega^2$ 
which we refer to by the name ``radion".
 The bulk scalar field induces the energy-momentum
 tensor of the conventional 4-dimensional scalar field with 
 the effective potential which depends on the radion. 

We can also determine the dark radiation $\chi^{\mu}{}_{\nu}$ by eliminating 
$G^{\mu}{}_{\nu}(h)$ from Eqs.~(\ref{p:einstein}) and
(\ref{n:einstein-1}),
\begin{eqnarray}
{2\over \ell}\chi^\mu{}_{\nu}
	&=&-{1\over \Psi}\left[ \Psi^{|\mu}{}_{|\nu} 
     	- \delta^\mu_\nu \Psi^{|\alpha}{}_{|\alpha} 
     	+{3\over 2}{1 \over 1-\Psi} \left(
     	\Psi^{|\mu} \Psi_{|\nu} -{1\over 2}\delta^\mu_\nu
     	\Psi^{|\alpha} \Psi_{|\alpha} \right) \right] 
     	\nonumber\\
&&
     	+{\kappa^2 \over 2}(1-\Psi)\delta^\mu_\nu V
     	-\frac{\kappa^2}{\ell}\frac{1-\Psi}{\Psi}
     	\left[\overset{\oplus}{T}{}^{\mu}{}_{\nu}
     	+  \left(1-\Psi\right)
     	\overset{\ominus}{T}{}^{\mu}{}_{\nu}   \right] \ .
     	\label{chi2}
\end{eqnarray}
Due to the property $\chi^\mu{}_\mu =0$, we have
\begin{eqnarray}
\Box\Psi =
	\frac{\kappa^2}{3\ell}(1-\Psi)\left[
  	\overset{\oplus}{T}
  	+(1-\Psi)\overset{\ominus}{T}\right]
	-{1\over 2(1-\Psi)} \Psi^{|\alpha} \Psi_{|\alpha} 
  	-{2\kappa^2\over 3}\Psi(1-\Psi)V   
  	\label{QST2}\ .
\end{eqnarray}
Note that Eqs.~(\ref{QST1}) and (\ref{QST2}) are derived from 
a scalar-tensor type theory coupled to the  additional scalar field.

Similarly, the equations for the scalar field on branes become
\begin{eqnarray}
&&\Box_h\eta-{V^\prime\over 2}+{2\over\ell}C
	=0  \ ,
	\label{p:KG}\\
&&\Box_f\eta-{V^\prime\over 2}+{2\over\ell}{C\over\Omega^4}
	=0 \ ,
	\label{n:KG}
\end{eqnarray}
where the subscripts refer to the induced metric on each brane. 
Notice that the scalar field takes the same value for both branes 
at this order.
Eliminating the dark source $C$ from these Eqs.~(\ref{p:KG}) 
and (\ref{n:KG}), we find the equation for the scalar field 
takes the form
\begin{eqnarray}
\Box_h\eta-V'_{\rm eff}
	=-\frac{\Psi^{|\mu}}{\Psi}\eta_{|\mu} \ .
	\label{p:KG-2}
\end{eqnarray}
Notice that the radion acts as a source for $\eta$. And we can also 
get the dark source as
\begin{eqnarray}
{2\over \ell}C=-{V^\prime\over 2}(1-\Psi)
	+{\Psi^{|\mu}\over\Psi}\eta_{|\mu}    \ .
	\label{C2}
\end{eqnarray}

Now the effective action for the positive tension brane which
gives Eqs.~(\ref{QST1}), (\ref{QST2}) and (\ref{p:KG-2}) can be read off as
\begin{eqnarray}
S&=&{\ell\over 2\kappa^2}\int d^4x\sqrt{-h}\left[
	\Psi R
	-\frac{3}{2(1-\Psi)}\Psi^{|\alpha} \Psi_{|\alpha} 
	-\kappa^2\Psi\left(\eta^{|\alpha}\eta_{|\alpha}
	+2V_{\rm eff}\right)
	\right]  
	\nonumber\\
&&
	+\int d^4x\sqrt{-h}~
	\overset{\oplus}{\cal L}
	+\int d^4x\sqrt{-h}~(1-\Psi)^2
	\overset{\ominus}{\cal L}  \ ,
\end{eqnarray}
where the last two terms represent actions for the matter on each brane.  
Thus, we found the radion field couples with the induced metric and 
the induced scalar field on the brane non-trivially. 
 Surprisingly, at this order, the nonlocality of 
 $\chi_{\mu\nu}$ and $C$
 are eliminated by the radion. 

 As this is a closed system, we can analyze a primordial spectrum
 to predict the  cosmic background fluctuation spectrum~\cite{soda2}.
  Interestingly, $\chi^{\mu}_{\nu}$ and $C$ vanishes in the single brane 
limit, $\Psi \rightarrow 1$, as can be seen from (\ref{chi2}) and (\ref{C2}).
 The dynamics is simply governed by Einstein theory with the single scalar 
 field. Therefore, we can conclude that the bulk inflaton can drive inflation
 when the slow role conditions are satisfied. 

\section{KK corrections }

 It would be important to take into account the KK effects
 as corrections to the leading order result. 
 Using our approach, in the single brane limit,
 we can deduce the effective action  with KK corrections as~\cite{kanno2}
 ( see also  \cite{Kanno} for recent developments)
\begin{eqnarray}
S&=&\frac{\ell}{2\kappa^2}\int d^4x\sqrt{-h}\left[
	\left(1+\frac{\ell^2}{12}\kappa^2V\right)R
	-\kappa^2\left(
	1+\frac{\ell^2}{12}\kappa^2V-\frac{\ell^2}{4}V''\right)
	\eta^{|\alpha}\eta_{|\alpha}
	-2\kappa^2V_{\rm eff}
	\right.\nonumber\\
&&\left.\hspace{3cm}
	-\frac{\ell^2}{4}\left(
	R^{\alpha\beta}R_{\alpha\beta}
	-\frac{1}{3}R^2
	\right)\right]
	+\int d^4 x \sqrt{-h}~ 
	{\cal L}_{\rm matter}
	+S_{\rm CFT} \ ,
\end{eqnarray}
where the effective potential at this order is defined by 
\begin{eqnarray}
V_{\rm eff}=\frac{1}{2}V
	+\frac{\ell^2\kappa^2}{48}V^2
	-\frac{\ell^2}{64}V^{\prime 2} 
	\label{2:effpt}\ .
\end{eqnarray}
and the last term comes from 
 the energy-momentum tensor of CFT matter $\tau_{\mu\nu}$.

\section{Conclusion}

 We derived the non-linear low energy effective action for the 
 dilatonic braneworld.
 We considered the bulk scalar field  with a nontrivial potential.
 Then, the constant of integration is determined completely.
 As a result, the effective theory reduces to the scalar-tensor
 theory with the non-trivial coupling between the radion and the bulk scalar
 field.  It turns out that $\chi_{\mu\nu}$ and $C$ becomes zero 
 when two branes get separated  infinitely. 
 This implies that the bulk inflaton can drive the inflation on the brane
 as far as the slow role conditions are satisfied. We also obtained
 KK corrections in the single brane limit. 

\vskip 1cm
\noindent{\em Acknowledgments}\\
\smallskip
This work was supported in part by  Grant-in-Aid for  Scientific
Research Fund of the Ministry of Education, Science and Culture of Japan 
 No. 155476 (SK) and  No.14540258 (JS).


\begin{thebibliography}{99}


\bibitem{RS1}
L.~Randall and R.~Sundrum,
Phys.\ Rev.\ Lett.\  {\bf 83}, 3370 (1999)
[arXiv:hep-ph/9905221].

\bibitem{bulk}
S.~Kobayashi, K.~Koyama and J.~Soda,
Phys.\ Lett.\ B {\bf 501}, 157 (2001)
[arXiv:hep-th/0009160];
Y.~Himemoto and M.~Sasaki,
Phys.\ Rev.\ D {\bf 63}, 044015 (2001)
[arXiv:gr-qc/0010035].

\bibitem{kanno1}
S.~Kanno and J.~Soda,
Phys.\ Rev.\ D {\bf 66}, 043526 (2002)
[arXiv:hep-th/0205188];
S.~Kanno and J.~Soda,
Phys.\ Rev.\ D {\bf 66}, 083506 (2002)
[arXiv:hep-th/0207029].


\bibitem{soda2}
J.~Soda and S.~Kanno,
Astrophysics and Space Science, {\bf 283}, 639 (2003), 
arXiv:gr-qc/0209086;
S.~Kanno, M.~Sasaki and J.~Soda,
Prog. Theor. Phys. {\bf 109}, 357 (2003), 
arXiv:hep-th/0210250.

\bibitem{kanno2}
S.~Kanno and J.~Soda,
arXiv:hep-th/0303203.

\bibitem{Kanno}
S.~Kanno and J.~Soda,
arXiv:hep-th/0312106.


\end{thebibliography}
\end{document}